\documentclass[aps,showkeys,showpacs]{revtex4}
\usepackage{epsfig}

\newcommand{\bc}{\begin{center}}
\newcommand{\ec}{\end{center}}
\newcommand{\bmi}{\begin{minipage}}
\newcommand{\emi}{\end{minipage}}
\newcommand{\bdi}{\begin{displaymath}}
\newcommand{\edi}{\end{displaymath}}
\newcommand{\bit}{\begin{itemize}}
\newcommand{\eit}{\end{itemize}}

\newcommand{\Ne}{N\'{e}el}

\newcommand{\bi}{\begin{itemize}}
\newcommand{\ei}{\end{itemize}}
\newcommand{\be}{\begin{equation}}
\newcommand{\ee}{\end{equation}}

\newcommand{\bea}{\begin{eqnarray}}
\newcommand{\eea}{\end{eqnarray}}   

\newcommand{\bdm}{\begin{displaymath}}
\newcommand{\edm}{\end{displaymath}}
\newcommand{\beas}{\begin{eqnarray*}} 
\newcommand{\eeas}{\end{eqnarray*}}
\newcommand{\bite}{\begin{itemize}}
\newcommand{\enite}{\end{itemize}}
\newcommand{\ben}{\begin{enumerate}}
\newcommand{\een}{\end{enumerate}}

\newcommand{\kag}{{\it kagom{\'e}{\ }}}

\begin{document}

\title{Quantum disorder due to singlet formation: The
Plaquette lattice}

\author{A.~Voigt}

\affiliation{Center for Simulational Physics,
Department of Physics and Astronomy,
University of Georgia, Athens GA 30605 USA
\footnote{Tel. +1-706-542-3867, Fax +1-706-542-2492,
E-mail: andreas@physast.uga.edu}}

\begin{abstract} 
I study the order/disorder transition due to singlet formation in a quantum
spin system by means of exact diagonalization. The systems is build by spin
1/2 on a two-dimensional square lattice with two different kinds of
antiferromagnetic Heisenberg interactions. The interaction $J_p$ connects 4
nearest neighbor spins on a plaquette. The interaction $J_n$ connects the
plaquettes with each other. If $J_p=J_n$ the systems reduces to the simple
square lattice case. If one of the interactions becomes sufficiently larger
then the other the purely quantum effect of singlet formation drives the
system into a disordered phase with only short range correlations in the
plaquettes and a spin gap. I study the transition point by evaluating the
spin gap and spin--spin correlations. I compare the results with previously
calculated data from a non--linear $\sigma$ model approach, spin wave theory
and series expansion calculations. I confirm a critical value of
$J_n\approx0.6$ for the quantum phase transition point.
\end{abstract}



\maketitle

\section{Introduction}

Low dimensional quantum antiferromagnets have attracted a great deal of
attention during the last decade, both theoretically and experimentally.
Such magnets show a wide variety of magnetic low--temperature behavior, like
magnetic long range order or spin disorder with or without a spin gap.
Especially the question how magnetic long range order is influenced by
different parameters like frustration, temperature, spin quantum number or
lattice topology has been addressed.

A wide variety of different numerical methods like quantum Monte Carlo,
renormalization group calculations, series expansion or spin wave theory has
been developed and used for carrying out research in this field. All these
methods have their advantages but also some disadvantages. Monte Carlo
calculations suffer severe from the sign problem in frustrated systems.
Series expansions and spin wave calculations may break down at phase
transition points. The most promising renormalization group theory, the
density matrix renormalization group is still limited to one--dimensional
problems. 

In this respect the exact diagonalization of finite lattices is still a very
powerful method because the calculated data is {\it a priori} exact for the
finite lattice under consideration. The exponentially fast growing Hilbert
space necessary for the calculations limits the enumeration for spin half
system to a total of N=36 spins yet. Still this task is computationally very
demanding and one of the computational challenges of the new century will be
the ongoing task to push this limit as far as possible by using all the
available resources at hand. Exact diagonalization will always be one of the
forerunners of using all the promised computer power coming up in the future
and nevertheless it will be not enough to go too far away.

Despite the mentioned limitations to rather small finite lattices it has
been shown that with particular care one can draw important conclusions
about the ordering behavior of different low--dimensional quantum spin
systems, e.g. the frustrated ferrimagnet in 1d \cite{ivrischo98}, the {\kag}
lattice in 2d \cite{lecheminant97,sindzingre00,hida00}, or the Heisenberg
antiferromagnet on the body--centered cubic lattice in 3d \cite{betts98}.

In this paper I will examine a particular interesting quantum spin system
which shows a purely quantum phase transition by singlet formation. There is
an experimental study of a new compound $Na_5RbCu_4(AsO_4)_4Cl_2$ under way
\cite{clayhold01} which suggest that this system might be described by the
model I consider here. I also note that a similar model has been studied
previously with a non--linear $\sigma$-model approach and spin wave theory
\cite{koga99}, plaquette expansion \cite{koga991} and Ising series expansion
\cite{singh99}. At the end I will compare the results of all this methods 
with each other.

\section{The Model}

I consider the Heisenberg antiferromagnet on a two--dimensional square
lattice with two different kinds of antiferromagnetic spin exchange 
parameters ${J_p>0}$ and  {$J_n>0$}.
~
\be
\hspace*{1cm}
H= {J_p} \sum_{inter}{{\hat s}_i{\hat s}_j} +
   {J_n} \sum_{intra}{{\hat s}_i{\hat s}_j}
\ee

\begin{figure}[ht]
\centerline{
\epsfig{file=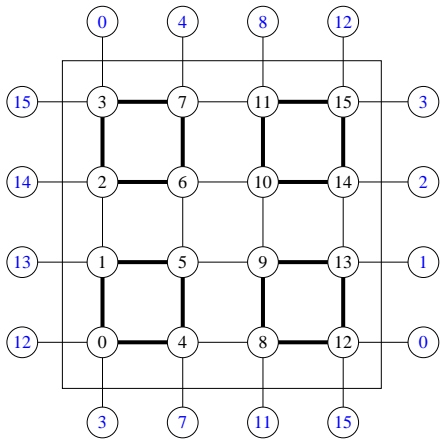,scale=0.4,angle=0}}
\caption{The Plaquette lattice with N=16 spins and periodic boundaries. 
Bold lines denote {$J_p$} bonds, thin lines denote {$J_n$} bonds. }
\label{latt_fig}
\end{figure}

The summation $\sum_{inter}$ with $J_p$ takes place between 4 nearest
neighbor spins on a particular plaquette. The $\sum_{intra}$ with $J_n$
takes place between nearest neighbors on different plaquettes (cf.
Fig.\ref{latt_fig}). The quantum spin ${\hat s}_i$ equals ${1\over2}$. For
simplicity I put {$J_p$=1} for the rest of the paper and take {$J_n$} as the
changing parameter.

I note that the Hamiltonian is symmetric if one exchanges ${J_n}$ and
${J_p}$ and therefore I consider the range of {$J_n \in [0,1]$ only. For
$J_n=1$ this system represents the simple square lattice.

\section{Results}
\subsection{The classical ground state}

The ground state of the classical model where the spins ${\hat s}_i$ can be
considered as classical vectors ${\vec s}_i$ is actually very simple.
Because there is no frustration in the system it will remain in the
\Ne--state for all values of $J_n>0$. This classical ground state is
illustrated in the left part of Fig.\ref{class}. Only if ${J_n}=0$ the
ground state becomes highly degenerated and then the \Ne--state is only one
of the possible ground states due to the free rotation of the spins in one
plaquette respectively to any other plaquette. This behavior is depicted in
the right part of Fig.\ref{class}.

\begin{figure}[ht]
\bc
\epsfig{file=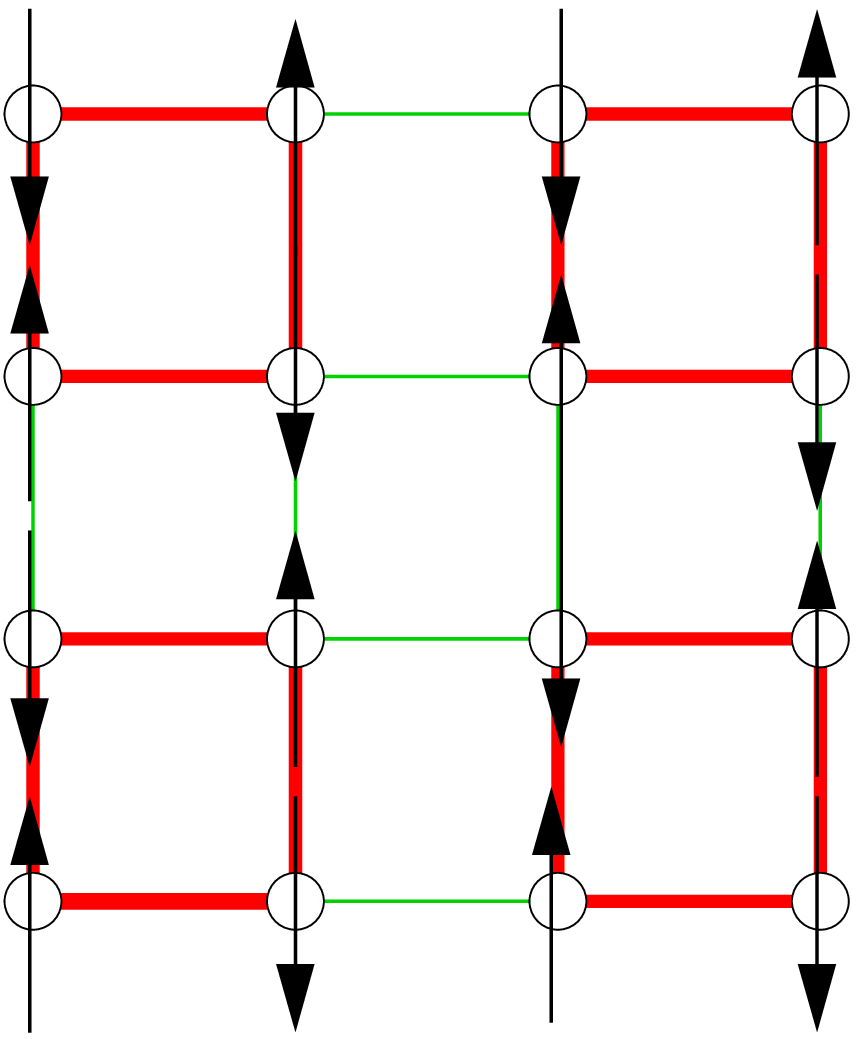,scale=0.4,angle=0}
\hspace{1cm}
\epsfig{file=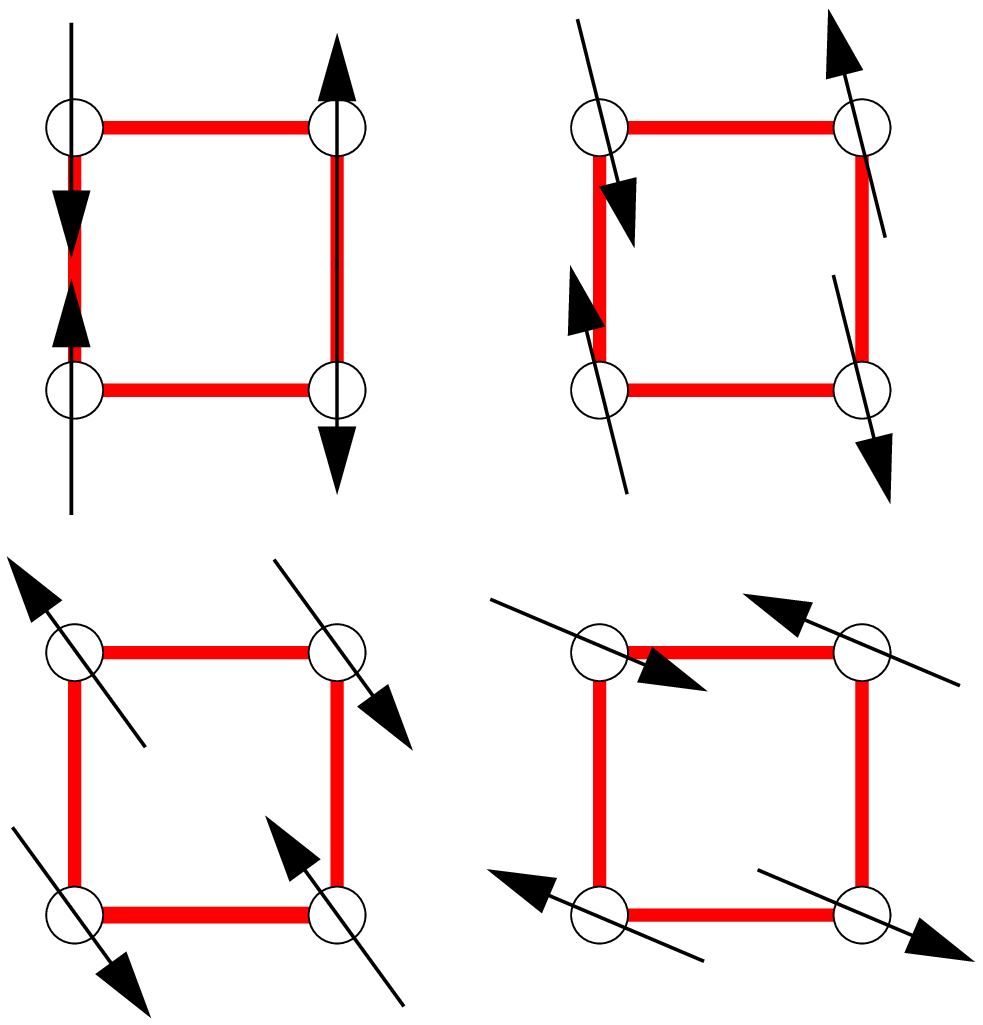,scale=0.4,angle=0}
\ec
\caption{The classical ground state spin configuration of the Plaquette
lattice. Left: The \Ne--state for ${J_n}>0$, right: ${J_n}=0$ with a free
rotation of the local axis in each plaquette.}
\label{class}
\end{figure}

Of course the question is now, what will happen if quantum fluctuations 
come into play? I will discuss this in the next paragraph by considering
the spin gap and spin--spin correlation data for the quantum system on finite
lattices with N=16,20,32 and 36 spins and periodic boundary conditions.

Since the Heisenberg Hamiltonian commutes with the square of total spin
${\hat S}^2$, ${\hat S}=\sum_i {\hat s}_i$, each eigen state of $\hat H$
belongs to a certain subspace of the Hilbert space with fixed quantum number
$S$ of the total spin (${\hat S}^2=\hat S(\hat S+1)$).  To calculate the
ground state and the first excited state of $\hat H$ for finite lattices I
use the Lanczos technique.

\subsection{The spin gap $\Delta$}

First I will analyze the spin gap $\Delta$

\be
\Delta = E_{min}(S_{min}+1) - E_{min}(S_{min}) .
\ee

For bipartite antiferromagnets like the one studied here one can show that
the ground state belongs to the $S=0$ subspace and the first excitation is a
triplet with $S=1$. Considering our model one can immediately note, that for
${J_n}=0$ all the plaquettes will be in a singlet state with the total spin
of the plaquette being zero. The first excitation will then be a state were
one plaquette is in its triplet state and a simple analytical calculation
then shows: $\Delta=1$. As stated for ${J_n}=1$ one obtains the simple
square lattice case where a \Ne--like magnetic long range order without a
spin gap is well established \cite{manousakis91}. For some value of $J_n$
between 0 and 1 I will observe the opening of a spin gap at the phase
transition point.

In Fig.\ref{gap} I show the data from exact diagonalization calculations for
lattices with up to 36 spins and an extrapolation to the thermodynamic
limit.

\begin{figure}[ht]
\bc
\epsfig{file=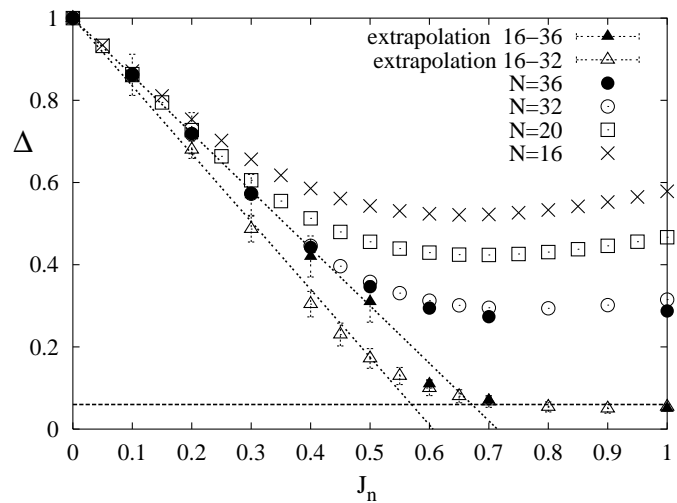,scale=0.4,angle=270}
\ec
\caption{ The spin gap $\Delta$ versus $J_n$. The crosses, squares and
circles represent the calculated data for N=16,20, 32 and 36 sites. The
triangles with the errorbars are the extrapolated data with two input sets:
N=16-32 and N=16-36. For the dashed and dotted lines see text.}
\label{gap}
\end{figure}

I use a finite size extrapolation $\Delta \propto N^{-1}$ for the entire
range of calculated data but will keep in mind that this is a reasonable
good approximation for $J_n\ge0.6$ only (apparent in the small errorbars)
but much less appropriate for $J_n\le0.5$ (visible in larger errorbars). I
used two data sets with N=16-32 and N=16-36. They give very similar result
for $J_n\ge0.6$, but deviate apparently for $0.3<J_n<0.5$. But even so one
can clearly distinguish the two phases, one with a finite spin gap and one
without and hence I draw two dotted fit lines for $J_n\le0.5$ and one dashed
fit line for $J_n\ge0.6$.

The extrapolated data shows a small finite value for $\Delta$ at $J_n=1$ but
this is due to the rough approximation only. I assume a finite spin gap
opens up above any value greater than the vertical dashed line (cf.
Fig.\ref{gap}) in the thermodynamic limit. The crossing of the two dotted
fit lines with the dashed one can then be approximated as a value for the
phase transition point: ${{J_n}^{crit}} \approx 0.65 \pm 0.05 $.

\subsection{The spin--spin correlation} 

In order to further evaluate the phase transition point I study the
spin--spin correlation data plotted in Fig.\ref{corr}. Here I show the
spin--spin correlation between two spins connected by a $J_n$ bond $\langle
\hat s_i \hat s_j \rangle_{intra}$ and between two spins connected by a 
$J_p$ bond $\langle \hat s_i \hat s_j \rangle_{inter}$.

\begin{figure}[ht]
\bc
\epsfig{file=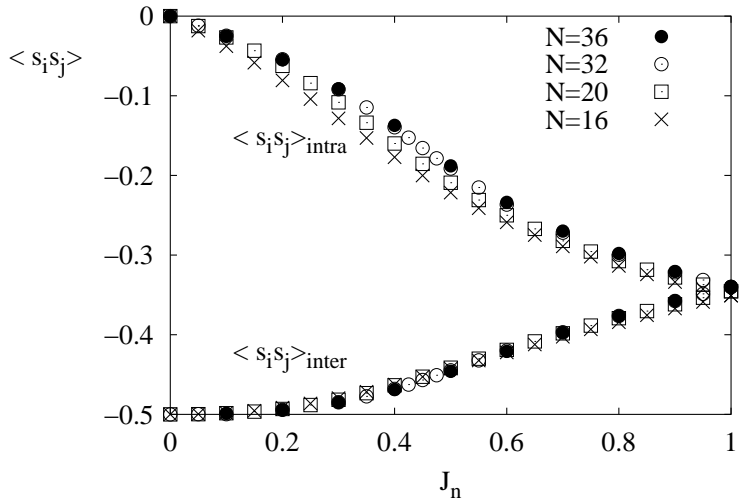,scale=0.4,angle=270}
\ec
\caption{
The spin--spin correlation $\langle \hat s_i \hat s_j \rangle$ 
between two spins connected by a $J_n$ bond (intra) or by a $J_p$ bond
(inter) versus $J_n$.}
\label{corr}
\end{figure}

The intra--plaquette correlation starts from 0 as expected for non--coupled
plaquettes and ends at a certain finite value for $J_n=1$. The
inter--plaquette correlation starts from $-1/2$ (this value can be
calculated analytically) and ends at the same value as $\langle \hat s_i
\hat s_j \rangle_{intra}$ due to the symmetry at this point.

Both curves show an inflection point at around $J_n\approx 0.5$ which can be
assumed to be connected to the phase transition. Therefore I carry
out a numerical differentiation $\partial{\langle \hat s_i \hat s_j
\rangle}_{intra} / \partial{J_n}$ in order to study this inflection point in
more detail in the left part of Fig.\ref{diffcorr}. By numerical
differentiation this inflection point is transformed into a minimum
$\delta_{min}$ which is then shown versus the system size in the right part
of Fig.\ref{diffcorr}.

\begin{figure}[ht]
\bc
\epsfig{file=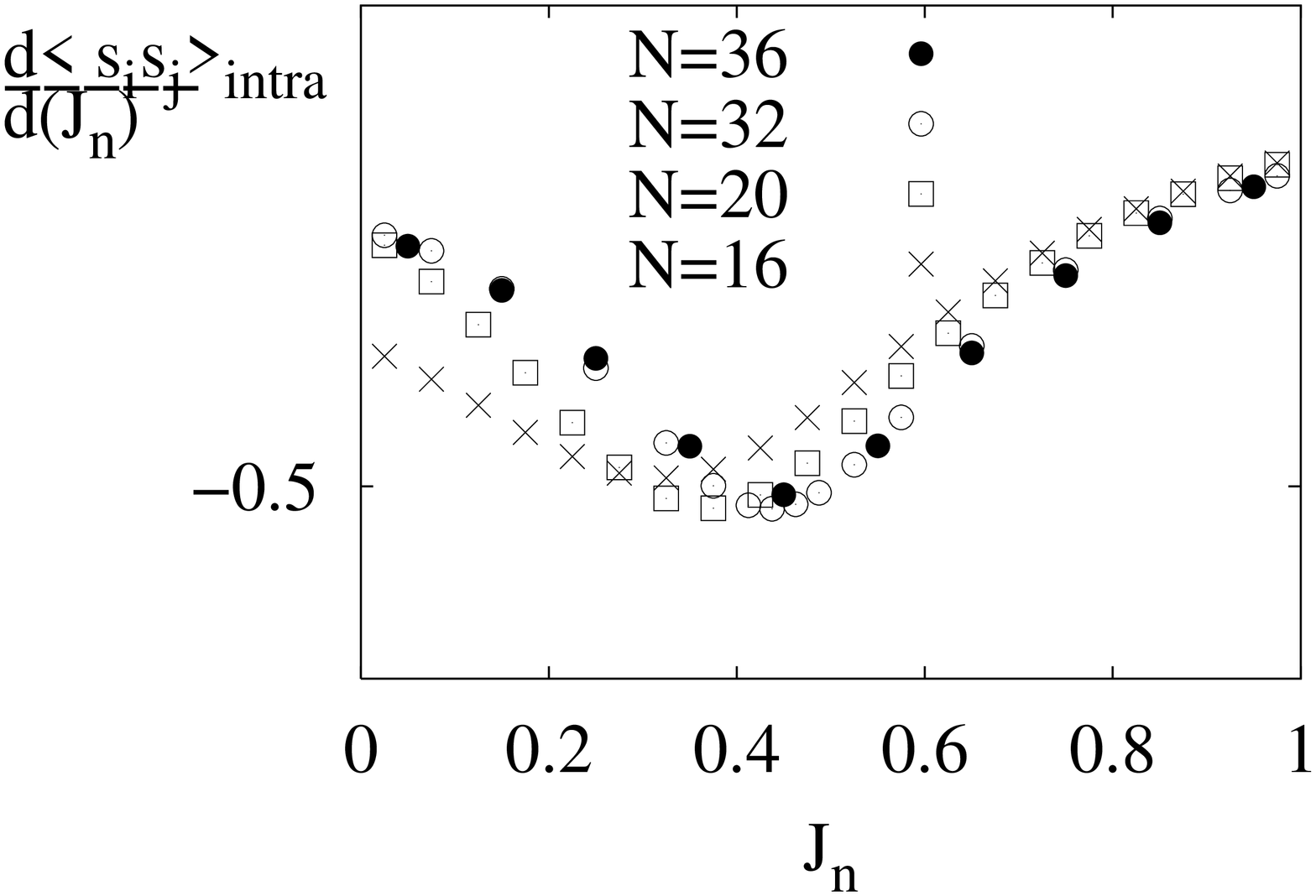,scale=0.25,angle=270}
\hspace{0.2cm}
\epsfig{file=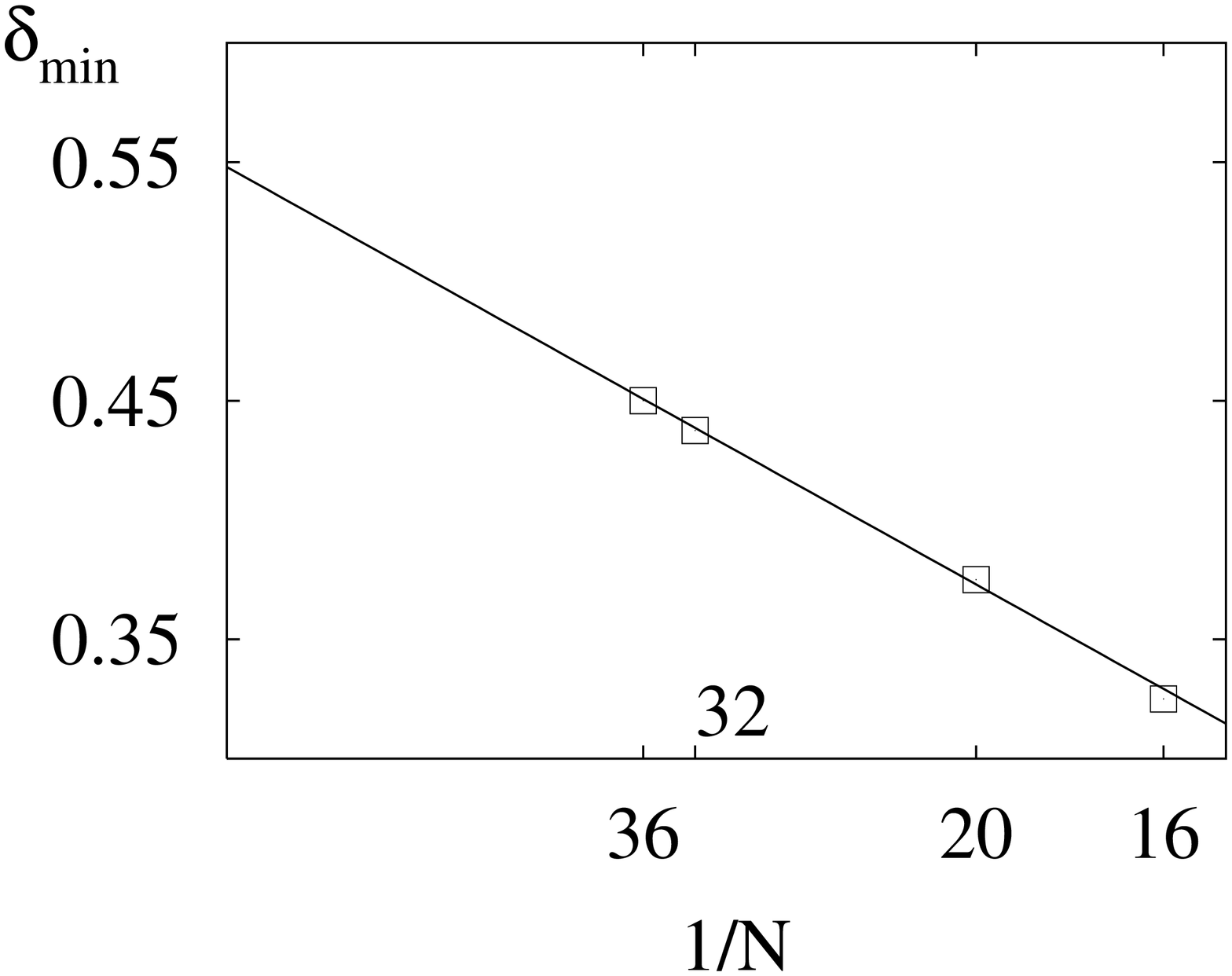,scale=0.25,angle=270}
\ec
\caption{
Left: The differential spin--spin correlation $\frac{ \partial \langle 
\hat s_i \hat s_j \rangle_{intra}} {\partial J_n}$ versus $J_n$. 
Right: The minimum position $\delta_{min}$ versus the inverse system 
size 1/N.}
\label{diffcorr}
\end{figure}

In the left part of Fig.\ref{diffcorr} one can see that with increasing
system size the minimum shifts towards greater $J_n$ and also gets deeper.
One may argue that in the thermodynamic limit one will observe a delta peak
right at the quantum phase transition point. In the right part of
Fig.\ref{diffcorr} I plot the peak position $\delta_{min}$ versus the
inverse system size and extrapolate a value $J_n^{crit}\approx0.55\pm0.05$.
This result is close to the value from the spin gap data.

\section{Conclusion}

In this paper I presented a exact diagonalization study of an
antiferromagnetic Heisenberg quantum spin systems with s=1/2 with two kinds
of interactions, the Plaquette lattice. I calculated the spin gap
and the spin--spin correlation for finite lattices and obtained a quantum
phase transition at $J_n^{crit}\approx0.6\pm0.1$. This is a purely quantum
phase transition due to singlet formation which has no classical
counterpart. Our result compares well with 4th order plaquette expansion
result of ${{J_n}^{crit}} \approx 0.54$ \cite{koga991} or the Ising series
expansion result ${{J_n}^{crit}} \approx 0.55$ \cite{singh99}. Our result
shows clearly that the spin wave result ${{J_n}^{crit}} \approx 0.112$ 
\cite{koga99} can be ruled out. The non--linear $\sigma$ model approach
\cite{koga99} does completely fail to predict any transition to an ordered 
phase without a spin gap. 

Whether this phase transition is first or second order is still to be
answered, the spin--spin correlation data with a developing delta peak
points more to a first order transition. But clearly a more detailed study
is necessary to clarify this and corresponding work is under way.

Anyway, this study has shown that the exact diagonalization of finite
lattices can be successfully used for complex many--body problems and that
the results may help to verify and extend the validity of other methods.


\begin{thebibliography}{10}
\expandafter\ifx\csname url\endcsname\relax
  \def\url#1{\texttt{#1}}\fi
\expandafter\ifx\csname urlprefix\endcsname\relax\def\urlprefix{URL }\fi

\bibitem{ivrischo98}
N.~Ivanov, J.~Richter, U.~Schollw\"ock, Phys. Rev. B 58 (1998) 14456.

\bibitem{lecheminant97}
P.~Lecheminant, B.~Bernu, C.~Lhuillier, L.~Pierre, P.~Sindzingre, Phys. Rev. B
  56~(5) (1997) 2521.

\bibitem{sindzingre00}
P.~Sindzingre, G.~Misguich, C.~Lhuillier, B.~Bernu, L.~Pierre, C.~Waldtmann,
  H.-U. Everts, Phys. Rev. Lett. 84~(13) (2000) 2953--6.

\bibitem{hida00}
K.~Hida, J. Phys. Soc. Jpn. 69 (2000) 311.

\bibitem{betts98}
D.~Betts, J.~Schulenburg, G.~Stewart, J.~Richter, J.~Flynn, J. Phys. A: Math.
  Gen. 31 (1998) 7685.

\bibitem{clayhold01}
J.~Clayhold, M.~Ulutagay-Kartin, S.-J. Hwu, M.-H. Whangbo, A.~Voigt, to be
  published .

\bibitem{koga99}
A.~Koga, S.~Kumada, N.~Kawakami, J. Phys. Soc. Jpn. 68~(2) (1999) 642.

\bibitem{koga991}
A.~Koga, S.~Kumada, N.~Kawakami, J. Phys. Soc. Jpn. 68~(7) (1999) 2373.

\bibitem{singh99}
R.~Singh, Z.~Weihong, C.~J. Hamer, J.~Oitmaa, Phys. Rev. B 60~(10) (1999) 7278.

\bibitem{manousakis91}
E.~Manousakis, Rev. Mod. Phys. 63~(1) (1991) 1.

\end{thebibliography}
\end{document}